# CO observations of SNR IC 443

Zhiyu Zhang[1, 2] Yu Gao[1] Junzhi Wang[3]

[1]Purple Mountain Observatory, CAS, Nanjing 210008
[2]Graduated School, CAS, Beijing 100080
[3]Department of Astronomy, Nanjing University, Nanjing 210093

**Abstract:** We present our $^{12}$CO and $^{13}$CO mapping observations of SNR IC 443, interacting with molecular clouds. It is the first large scale high resolution $^{13}$CO mapping observation in the surrounding region. The morphologies of IC 443 in $^{12}$CO and $^{13}$CO are compared with the optical, infrared, Spitzer far-infrared, X-ray, and neutral atomic gas (HI). We also make comparison and analysis in the kinemics, using the date-cubes of $^{12}$CO, $^{13}$CO and HI, to help distinguish the complicated gas motions in the shocked regions. Based on the work of Wang & Scoville (1992), we present a new model to explain the origin of coexistence of shocks with different speeds in a rather small region at the central clump B. We test this new model via analyzing the HI and CO distributions in both velocity and space domains. We also establish the relationship between the dissociation rate of the shocked molecular gas and the shock velocity in this region. Finally, we derive the optical depth of $^{12}$CO with the $^{13}$CO spectra in clump B, and discuss the validity of the assumption of optically thin emission for the shocked $^{12}$CO.

Massive stars are born in molecular cloud cores, and evolve into expanding HII regions. Supernova explosion occurs when they evolve into their final destination signifying the death of the massive stars. The energetic explosion interacts with the ambient interstellar medium (ISM), spreading their 'ashes' into the surroundings, and thus depositing a huge amount of heavy elements in ISM for the star formation in the next generation. When the supernova remnants (SNRs) interact with ambient molecular clouds, the shocks can compress, heat up, ionize, and dissociate the molecular medium. All these contribute to the destruction of viral equilibrium from turbulence and further trigger or suppress the star-forming activity in the surrounding molecular clouds[3][6][17]. So, the feedbacks of SNRs to their parent molecular clouds play an important role in all stages of star formation.

The typical scale of molecular clouds is about 50-100 pc, whereas the scale of HII regions created by massive stars of 8-12 $M_\odot$ is normally less than 10 pc. Thus, the SNRs born from low mass end of massive stars are likely to interact directly with surrounding molecular clouds[7]. At a distance of 1.5 kpc (Fesen 1984)[13] and on the opposite of the direction of the Galactic centre, IC 443 (G189.1 + 3.0)is one of the first and best-studied unique case of SNRs that are interacting with nearby molecular clouds[16]. IC 443, close to the HII region S249, is located in an active star-forming region which is associated with the Gem OB1 association[13]. Moreover, being in the radiative phase, IC 443presents strong emission in optical[10], X-ray[2], radio continuum[17], cosmic ray[1], and infrared[20]. In optical, x-ray, and radio continuum images, IC 443 is separated into two semi-circular shells with a radius of 6pc -in the northeast (NE), and 10 pc -in the southwest (SW). Judging from the estimated expanding speed, it has an age of ~$10^4$ years.

The shell of IC 443 is expanding at a speed of around 100 km s$^{-1}$, and is interacting strongly with ambient molecular clouds, atomic gas (HI gas), and other ISM[4]. Quite a few observations have been made of the shock tracers including HI gas, molecular hydrogen ($H_2$) pure rotational line, forbidden line of 63μm [OI][21], and $H_2O$ infrared lines[27]. All these observations show that a single velocity shock model can not explain the presence of different kinds of shock tracers.

The diameter of IC 443 is about 40′ (~15 pc), which is on a rather large scale, so the kinematics of different clumps can not be expected to be uniform. As a result, the comparison of the quiescent gas with shocked gas in both spatial and velocity domains can give precise information that distinguishes the shocked components from quiescent ones. A new $^{12}CO$ and $^{13}CO$ mapping observation on a large scale will help study the shock feedback to molecular clouds in IC 443.

Although various CO mapping observations of IC 443 have been reported, most of them are either sparse large scale mapping with low angular resolution[24][11], or high resolution interferometer studies in individual clumps[30][31]. Whereas the optically thin $^{13}CO$ lines have only been observed in several single points[30][31][8], and our observation is the first systematic $^{13}CO$ mapping observation of IC 443. Our simultaneous $^{12}CO$ mapping also gives better spatial sampling than Dickman et al. (1992, hereafter D92)[11].

This paper is organized in the following way. The first section introduces observations and data reductions; the second section is the multi-band imaging comparison in IC 443; the analysis of the channel maps of $^{12}CO$, $^{13}CO$ and HI 21cm data-cubes, and the kinematical characteristics of the shocked molecular clouds are presented in the third section. The fourth section shows a new simple model in clump

B that can naturally explain why there are multiple shocks with different velocities in compact regions and the relation between the dissociation rate and the shock speed as well; in the fifth section, the optical depth of shocked $^{12}$CO in Region B is given by $^{13}$CO spectra, and the validity of the assumption of optically thin emission for the shocked $^{12}$CO is discussed. Conclusions are summarized in Section six.

## 1. Simultaneous $^{12}$CO (1-0), $^{13}$CO (1-0) and C$^{18}$O (1-0) observations of SNR IC 443

We performed the mapping observations of the $^{12}$CO J=1-0, $^{13}$CO J=1-0 and C$^{18}$O J=1-0 lines simultaneously in IC443 in October 2006 and March 2007, using the 13.7 m millimeter telescope of the Purple Mountain Observatory (PMO), located in Delingha, China. The typical system temperature of the 3mm SIS mixer receiver is 170-220 K, and the main beam efficiency $\eta^*_{mb}$ is about 0.67 at 110GHz with the main beam size of 50″ x 54″. Three 1024 channel acoustic-optic spectrometer (AOS) backends were used, and the velocity resolutions for $^{12}$CO, $^{13}$CO and C$^{18}$O are 0.37 km s$^{-1}$, 0.11 km s$^{-1}$, and 0.12 km s$^{-1}$ respectively. The integration time per observed point is typically from 1 minute to 5 minutes, resulting in a RMS noise level of 0.1-0.3 K at a velocity resolution of ~0.2 km s$^{-1}$.

With position-switching mode, we used $\alpha(2000) = 06^h17^m53^s.77$, $\delta(2000) = +22°31'41"$, located in the emission-free region in the center of IC 443, as the sky reference point, which is much closer to most molecular clumps than that used in D92, and thus better baselines were obtained. To save the observation time, we mapped the entire region of ~45′ x 40′ with a step of 2′ in the first stage, and then made a one-beam-sampling (with a step of 1′) mapping observations in the strong emission regions of ~35′ x 35′. In the peaks of $^{12}$CO, especially in clump B and G, we made deep integration for several hours.

All spectral data are reduced with GILDAS package, and antenna elevation calibration is also made. We make linear baseline subtractions to most spectra; order 2 or 3 baseline subtractions are used for spectra with non-flat, curved baselines (~5%); as for spectra with obvious standing waves(~5%), we use Fourier filter technique to get rid of the high frequency standing wave components[15]; bad scans are dropped(~10%). The data-cubes of $^{12}$CO and $^{13}$CO are made with XY_MAP package in GILDAS, in several velocity channel widths at a spatial resolution of 1′. IDL, MIRIAD and KARMA are used in the analysis process.

We make a multi-band comparison of our $^{12}$CO and $^{13}$CO results with the archival data of Spitzer MIPS 24 μm, 70 μm and 160 μm images, optical R-band archival image from DSS, 2MASS archival image in Ks band, ROSAT X-ray archival data, VLA 21 cm HI/continuum archival data. DSS, 2MASS and ROSAT data are downloaded from skyview website.

Spitzer MIPS images are archival post-BCD data, which were observed in October 2004, with the observation numbers of r4616960, r4617216 and r4617472 (PI: George Rieke). We used the median value of these three scans for both 24 μm and 70 μm images. While for 160 μm data, we used IMBIN in MIRIAD to combine every nearby 4 pixels to present the morphology, and avoid the influence from missing scans.

VLA HI 21cm spectral data come from Lee et al. (2008)[17]. This HI mapping was observed in October 2002 with Arecibo and the VLA ( the project number: AK0537, PI: koo, Bon_Chul) . This observation was made with mosaic mode to cover a large area of ~1°, in D array configuration. The missing flux was well calibrated by with HI observation of Arecibo. The final synthesis beam is about 40″, close to our CO resolution.

## 2. Multi-band images comparison of SNR IC 443

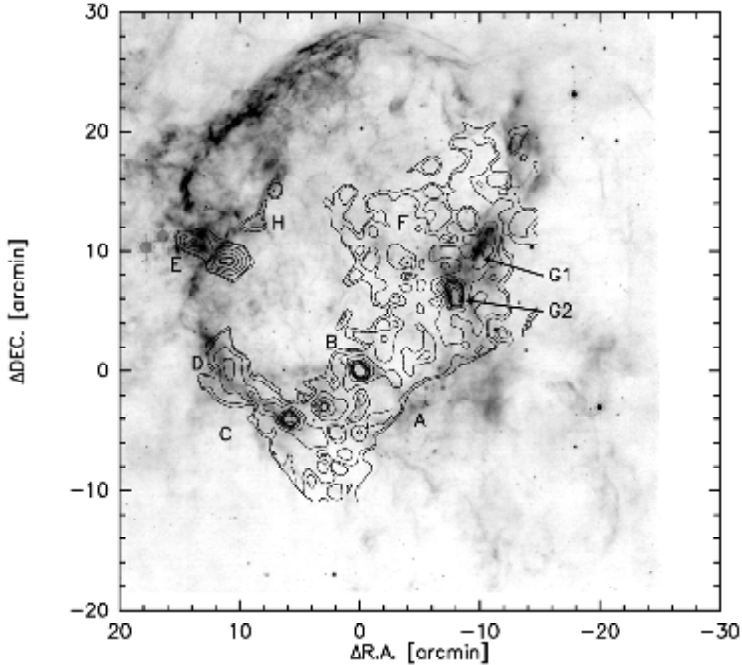

**Fig. 1.** Spitzer 24 μm image with $^{12}$CO integrated intensity contours overlaid. The contours are based on the 1′ spacing map, and the levels are 6, 10, 15 to 40, with a step of 5 K km s$^{-1}$. The map coordination is offset in arcminute with respect to the map center ($\alpha(2000) = 06^h17^m16^s$, $\delta(2000) = +22°25'41''$). Clumps are denoted with the nomenclature in Dickman et al. (1992, hereafter D92), and the G clump is resolved into two components as G1 -- the strongest $^{13}$CO peak, and G2 – the original G in D92.

Fig 1 shows that $^{12}$CO is dispersively distributed in this whole SNR, and the quasi-shell structure in the southern half matches the Spitzer 24 μm image. Peaks of $^{12}$CO surround the SNR shell, displaying a few small clumps (B, C, G, etc. following the nomenclature in D92). G clump is resolved into two peaks as G1-- the strongest $^{13}$CO peak, and G2 – the strongest $^{12}$CO peak, with the separation of 4′ (~1.5 pc) in the distance of 1.5 kpc.

The distribution of $^{12}$CO clumps is very similar to that shown in D92, while our map shows more subtle structures with 2 times higher gain in the spatial sampling resolution in most regions.  Compared with Spitzer 24 μm image, $^{12}$CO emission is concentrated in the inner shell of the infrared shell structure. Whereas in the northeastern (NE) region that shows the strongest 24 μm emission, $^{12}$CO does not have detectable emission in principle. Clump G1, one of the strongest peaks in 24 μm emission, matches well the $^{13}$CO peak position.   In D92′s clump A, we do not find any $^{12}$CO emission excess compared to its local environment, but $HCO^+$ (J=1-0)[11] and 24 μm show strong emission here.

It is generally believed that the line emissions, i.e. $H_2$ 0-0 S(0) 28.2 μm, [Fe II] 26 μm and [S I] 25 μm, contribute most of the Spitzer 24 μm broad band emission, whereas the continuum emission from grains and dust can be neglected in most regions. These three lines are rotational spectrum of molecular gas, fine structure spectrum of atomic gas, and spectrum of ionized gas respectively. These emission mechanisms are absolutely different from the 24 μm mid-IR dust emission in star-forming regions[20]. As the Spitzer IRS spectra show[20] that [Fe II] 26 μm spectral emission dominates the 24 μm emission in the NE region, which is consistent with our result of the non-detection of CO (cold gas) in the NE region, while in clump G2 and clump B regions, all these three line emissions are important to 24 μm

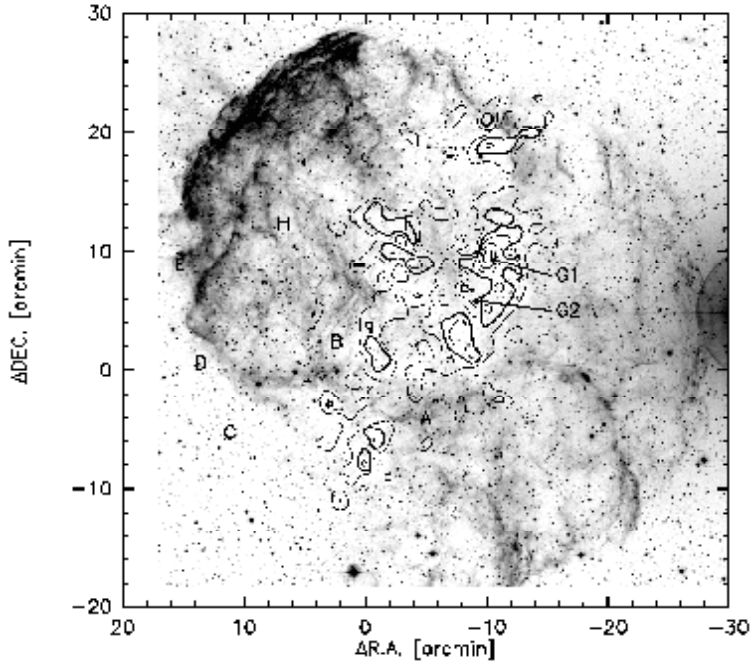

**Fig. 2.** Optical R-band image overlaid with $^{13}CO$ integrated intensity contours. Contour levels are 1, 3, 5, 7 K km s$^{-1}$. The map coordination is the same as in the Fig. 1.

Fig. 2 shows the anti-correlation between $^{13}CO$ distribution and optical R-band image. $^{13}CO$ is mainly distributed in the western part of the SNR, like a long tilted slice, separating the optical image into two semi-spheres: shell-like strong emission structure in the NE and weak filament in SW regions.

In IC 443, Hα line emission dominates the broad-band emission in R-band image[10]. Because the NE region is essentially free of molecular gas with only diffuse atomic gas detectable, the ionization and recombination processes after the shock dominate the cooling mechanism. The tilted molecular clouds, located mainly in the foreground of the SNR, surrounds the inner cavity region like a ring[4]. The optical extinction in G clump is AV ~3-4 mag[30]. So the anti-correlation between optical and $^{13}CO$ could be due to their different distributions in the line of sight. The shock induced ionization and recombination processes in low density atomic gas component dominate the cooling mechanism in SW and NE semi-sphere regions, so Hα emission there is stronger than in other regions.

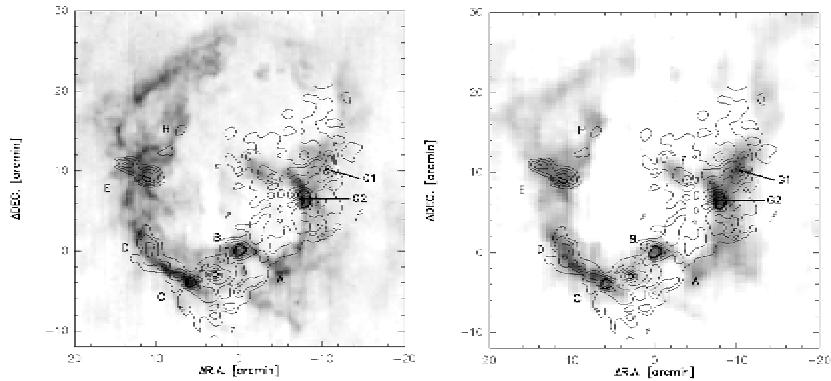

**Fig. 3.** The left panel shows the Spitzer 70 μm image overlaid with $^{12}$CO integrated intensity contour. The right panel shows the Spitzer 160 μm image overlaid with $^{12}$CO integrated intensity contour. The contour levels are the same as in Fig 1.

The most striking feature in Fig 3 is that both the peak distribution and the morphology of $^{12}$CO integrated intensity map match the Spitzer 70 μm (left) and 160 μm images (right) except for the NE regions. In the NE and SW regions, the morphology of 70 μm indicates good match with 1.4GHz radio continuum outer shell structure, while the 160 μm image shows only weak emission in NE and SW regions, with most emission in the south and west, where the SNR-molecular interaction plays the leading role.

Rho et al. (2001) observed several points in the shell structure with ISO LWS. Their results show that the Spitzer 70 μm emission is dominated by [O I] 63 μm. At the same time, the 160 μm emission, though shown as shell-like structure in overall appearance, is weaker in the NE regions and stronger in the shocked southern regions compared with 70 μm image. [CII] 158 μm emission is believed to dominate the Spitzer 160 μm band in IC 443[21]. The cooling mechanism in shocked gas is mainly [OI] and [CII] emission when shock speed is below 20 km s$^{-1}$[12], so we could infer that the interactions between shocks and molecular clouds in these regions are very strong so that the shock speed is much reduced.

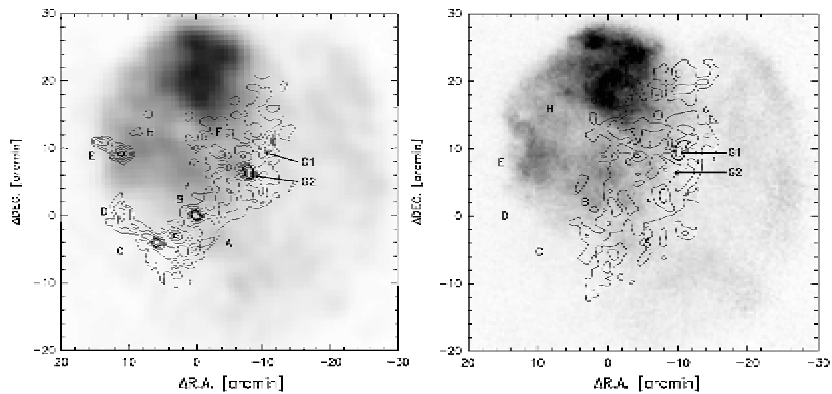

**Fig. 4.** The left panel shows the x-ray low energy (0.5 - 0.7keV) image of ROSAT-PSPC overlaid by $^{12}$CO integrated intensity contours. The right panel shows ROSAT-PSPC x-ray total energy intensity (0.1-2.2 keV) overlaid by $^{13}$CO integrated intensity contours. The contour levels are the same as in Fig 1.

Fig 4 shows the morphologies of soft X-ray compared with $^{12}$CO and $^{13}$CO distributions. The north region has the strongest X-ray emission, while only diffuse X-ray emission is located in the rest of the regions, outlining the entire SNR area in spherical structure. The hot gas concentration in X-ray emission is located exactly in the north region where warm and cold dust and gas are essentially absent, filling in the empty structures shown in 70, 160 μm images and in cold molecular gas. The ionized gas has a lower temperature in the western region than in the east, which can be inferred from X-ray morphology[2] on the large scale. The images of total intensity and high energy intensity are very similar, and both images show anti-correlation with the molecular gas.

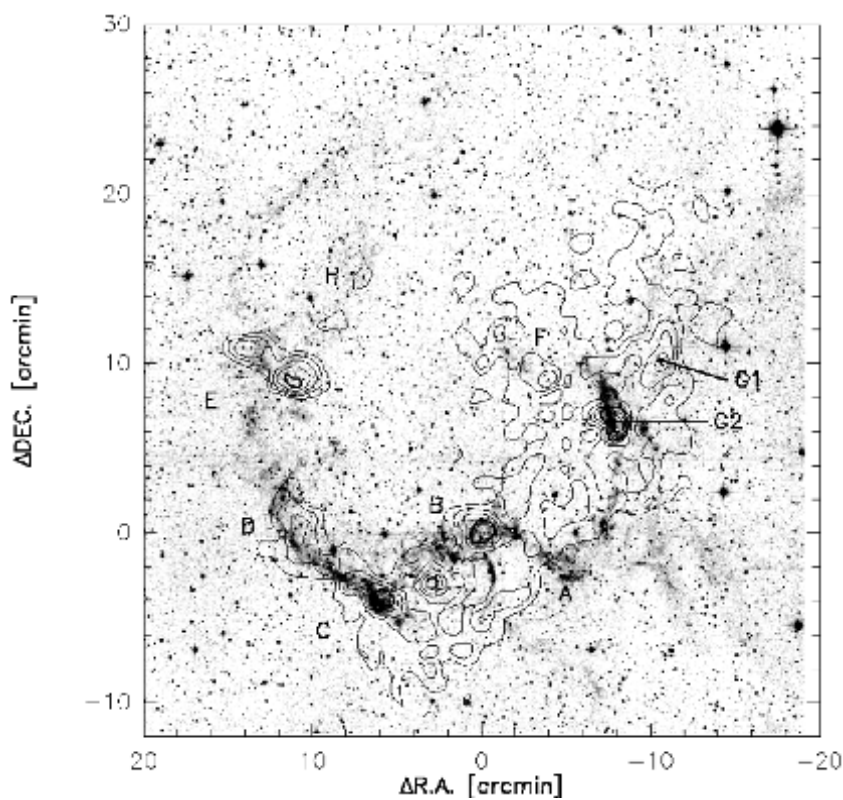

**Fig. 5.** 2MASS Ks band image overlaid with $^{12}$CO integrated intensity contours. The contour levels are the same as in Fig 1.

Fig 5 presents the distribution of 2MASS Ks near-infrared emission and $^{12}$CO integrated intensity. The 2.12 μm vibration-rotational transition of H$_2$ molecule dominates the 2MASS Ks band[21] and has a similar morphology with $^{12}$CO at most clumps, while in clump F, E, H, etc., 2MASS Ks-band emission is not obvious. In clump A, where strong 2MASS emission is presented, $^{12}$CO does not have any obvious emission peaks, neither does $^{13}$CO.

Rho et al.(2001) proposed that the atomic fine structure lines ([CI],[OI] etc.) are mainly excited by fast J-shocks with a speed around ~100kms$^{-1}$ in the NE regions, while the H$_2$ molecule emission, excited by slow C-shock with a speed of ~30kms$^{-1}$ is strong in the southern clumps. Both $^{12}$CO and H$_2$ molecule NIR emission can trace shocked molecular gas. Judging from the comparison in Fig. 5, the $^{12}$CO emission appears to better trace shocks with higher velocity and higher density than H$_2$ molecule emission.

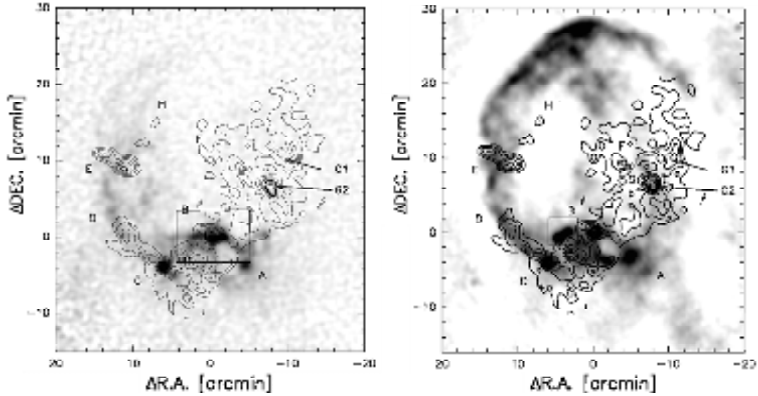

**Fig. 6.** The left panel shows the shocked HI gas (integrated flux in the velocity range from -160 to -20km s$^{-1}$) overlaid with $^{12}$CO integrated intensity contours. The contour levels are the same as in Fig 1. The box in the left panel presents the region selection of Fig 11. The right panel shows the 1.4 GHz radio continuum emission overlaid with $^{12}$CO integrated intensity contours. The two point sources in the small box around clump B are extragalactic sources.

The left panel in Fig 6 presents the shocked HI gas distribution, which is integrated from -160 km s$^{-1}$ to -20 km s$^{-1}$, with the assumption that most shocks have radial velocities in this range. Compared to the radio continuum, the extended shocked HI has a clear interface and peaks at clumps A, B and C. The shocked HI gas is stronger in the southern and eastern clumps than in the western ones, and only matches $^{12}$CO at clumps B and C, where the shock moves almost along the line of sight.

The right panel in Fig 6 presents the 1.4 GHz radio continuum overlaid by $^{12}$CO contours. Apart from the similar structure seen in the Hα optical image in the NE shell, the radio continuum also shows very strong emission at the interface between the shock and molecular clouds in the southern regions. Besides clumps F, H and G1, every $^{12}$CO peak has the counterpart of radio continuum. The shell structure of $^{12}$CO on large scale also matches radio continuum well.

## 3. The kinematics of $^{12}$CO, $^{13}$CO and HI 21cm gas

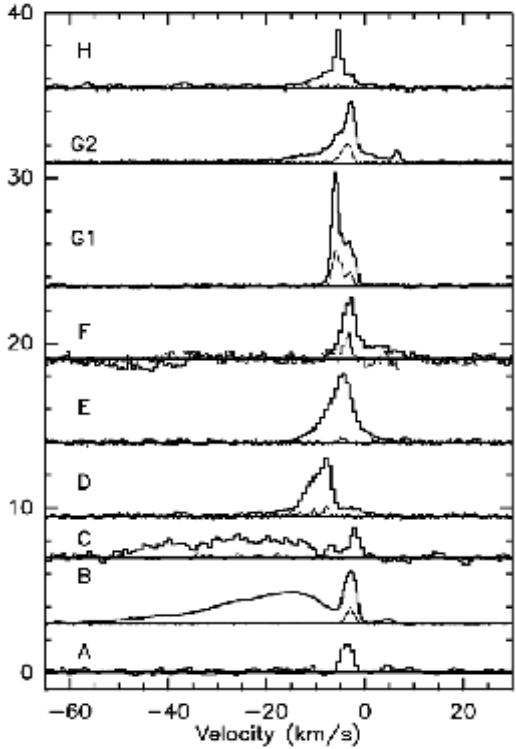

**Fig. 7.** The spectra in the central peak areas of the clumps. In order to get a better S/N, spectra in G2are the averaged result of a 3′ x 3′ region.   The bold lines are $^{12}$CO spectra, and thin lines are $^{13}$CO spectra.

Fig 7 presents the typical spectra of $^{12}$CO and $^{13}$CO in all clumps with shock interaction. Most $^{12}$CO spectra have broad velocity components with linewidths larger than those of normal molecular clouds (<10 km s$^{-1}$), and have multiple components in velocity. These broad $^{12}$CO line profiles deviate from Gaussian profiles, and do not have any $^{13}$CO counterparts which mostly trace the quiescent unshocked gas.

$^{13}$CO spectra with double peaks only appear in clump G1, whereas all $^{13}$CO spectra have single Gaussian peak with a linewidth less than 10 km s$^{-1}$ in the other clumps. The velocity variation between the line centers of different clumps reaches10 km s$^{-1}$. Only weak $^{13}$CO is detected in the clumps C, E and H, where the shock conduces to strong $^{12}$CO emission.

The shocked $^{12}$CO spectra have broad linewing profiles, so one can use the difference between the linewing and the quiescent components to find the shocked spectral components. While in IC 443, only the shocks in clumps B and C are propagating along the line of sight, whereas in other clumps only projected components can be observed. As a result, the line profiles of the shocked gas and quiescent gas are mixed. Moreover, the broad linewing components from shocked gas show absorption by the cold molecular clouds in the foreground in some clumps[30][31][33][27]. The information derived only from $^{12}$CO is not precise enough to distinguish the shocked components from environmental gas. Whereas $^{13}$CO, as an excellent molecular gas tracer, is optically thin in most cases and its line profile is less affected by the shock than $^{12}$CO. It is better to use the $^{13}$CO as the quiescent gas tracer to help find the shocked $^{12}$CO components.

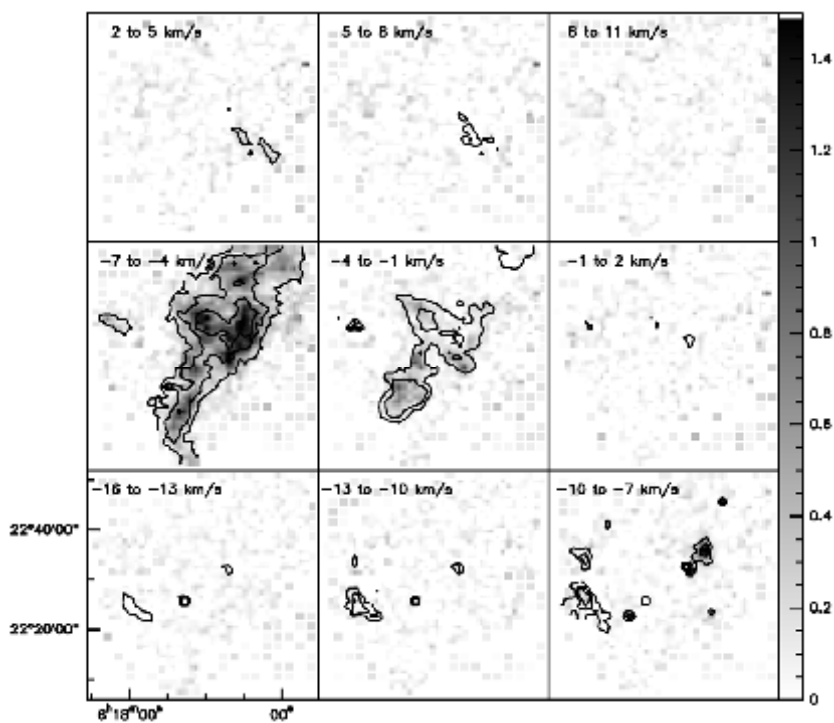

**Fig. 8.** Channel maps of $^{12}$CO and $^{13}$CO in the velocity range from -16 km s$^{-1}$ to 11 km s$^{-1}$. The contours are $^{12}$CO integrated intensities, with contour levels of 2, 4, 6 K km s$^{-1}$. The gray scales are $^{13}$CO integrated intensities in each panel.

Fig 8 presents the channel maps of $^{12}$CO J=1-0 and $^{13}$CO J=1-0 in the whole observed regions. The $^{13}$CO emission is concentrated in the velocity range from -7 km s$^{-1}$ to -1 km s$^{-1}$. There is a clear velocity gradient from SW to NE, which indicates that the quiescent molecular clouds are rotating. In the velocity range from -10 km s$^{-1}$ to -7 km s$^{-1}$, strong $^{13}$CO is only detected in a compact region of clump G1. Weak $^{13}$CO is detected in the SE region in a narrow velocity range, which presents the environmental molecular gas distribution.

$^{12}$CO emission presents very broad velocity coverage ($>$20 km s$^{-1}$), especially in the southern clumps with blue-shifted speed. In the velocity range from 2 kms$^{-1}$ to 8 kms$^{-1}$, there is a strip-like $^{12}$CO component which has weak $^{13}$CO counterpart. The average ratio of I$^{12}$CO/I$^{13}$CO in this region is 7. So we infer that this cloud could be in the foreground or background of the main clouds. Similarly, the narrow emission peak ( from -10 km s$^{-1}$ to -7 km s$^{-1}$ )in clump H could be an independent molecular clump, free from the interaction between the SNR and main clouds.

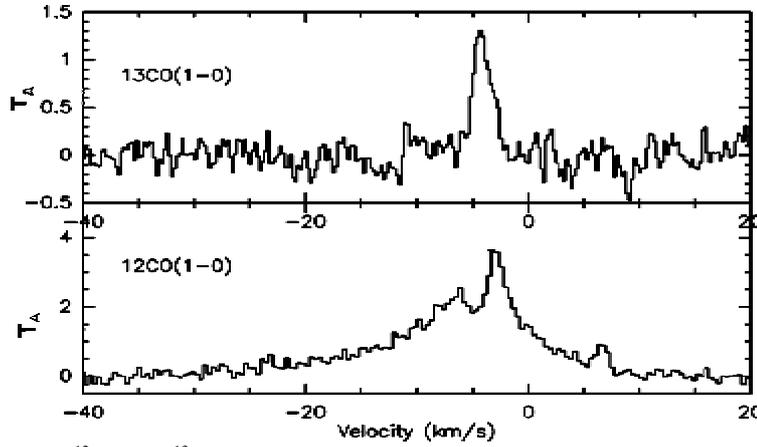

**Fig. 9.** $^{12}$CO and $^{13}$CO spectra in the central position of clump G2 (6 ′ in RA and 8′ in DEC in NE direction relative to the clump B ).

$^{12}$CO spectra in clump G2 have distinct double peaks line profile, which is caused by the absorption of the cold molecular cloud in the foreground of the SNR. As shown in Figs. 7 and 9, $^{13}$CO spectra have a line center speed corresponding to the center of the two peaks of $^{12}$CO. From this observational evidence, we infer that the pre-shocked environmental molecular gas is located just in front of the shock wave plane, which is consistent with the results found in the literature[30].

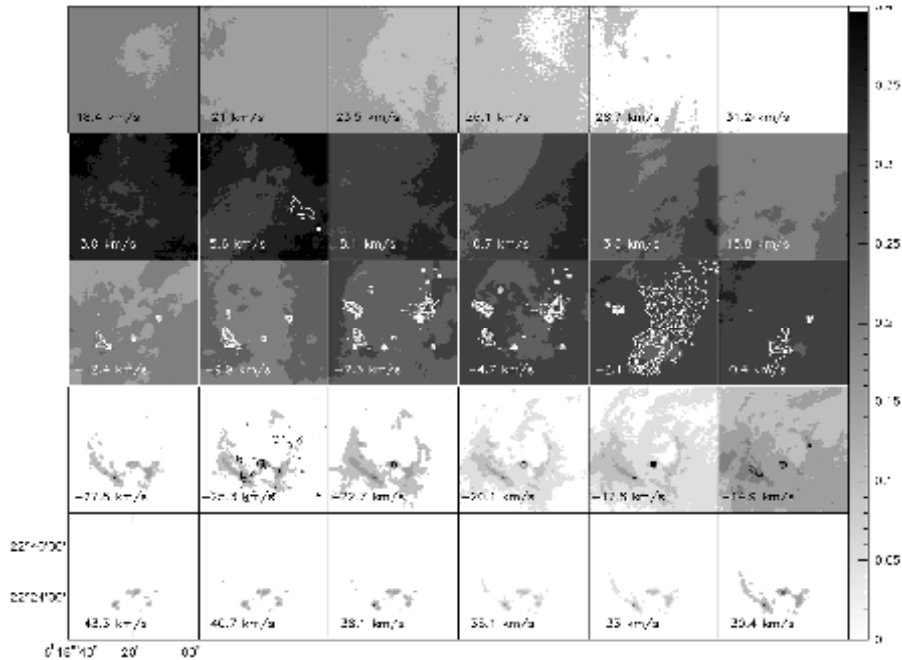

**Fig. 10.** Channel maps of $^{12}$CO and HI intensities in the velocity range from -43 km s$^{-1}$ to 31 km s$^{-1}$. $^{12}$CO contour levels are from 0.4 to 2.0 K with the interval of 0.4 K. Contours in the interval between -12.4 km s$^{-1}$ and 15.8 km s$^{-1}$ are plotted in white for high contrast to the background. The gray scales are HI intensities, whose

values are 0.075, 0.15, 0.3, 0.6, 0.9, 1.05, 1.2 Jy/beam km s$^{-1}$.

Fig 10 presents the channel maps of $^{12}$CO and HI gas in the velocity range from -43 km s$^{-1}$ to 31 km s$^{-1}$. HI gas is much more diffuse than molecular gas in all positions. In the velocity range from -10 km s$^{-1}$ to -2 km s$^{-1}$, HI gas presents extremely diffuse absorption feature toward the continuum in the background, showing the shell structures of the radio continuum.

The quiescent HI gas has a larger velocity coverage than that of $^{12}$CO, which emits mostly in the velocity range from -7 km s$^{-1}$ to 0 km s$^{-1}$, centering as -4 km s$^{-1}$. While the strongest HI gas distributes in a velocity range from 0 km s$^{-1}$ to 10 km s$^{-1}$, peaking at 5 km s$^{-1}$. This demonstrates that the quiescent HI gas and molecular gas systematically deviate in velocities.

## 4 The comparison between HI gas and $^{12}$CO in clump B

### 4.1. The comparison of HI and $^{12}$CO in velocity channel maps

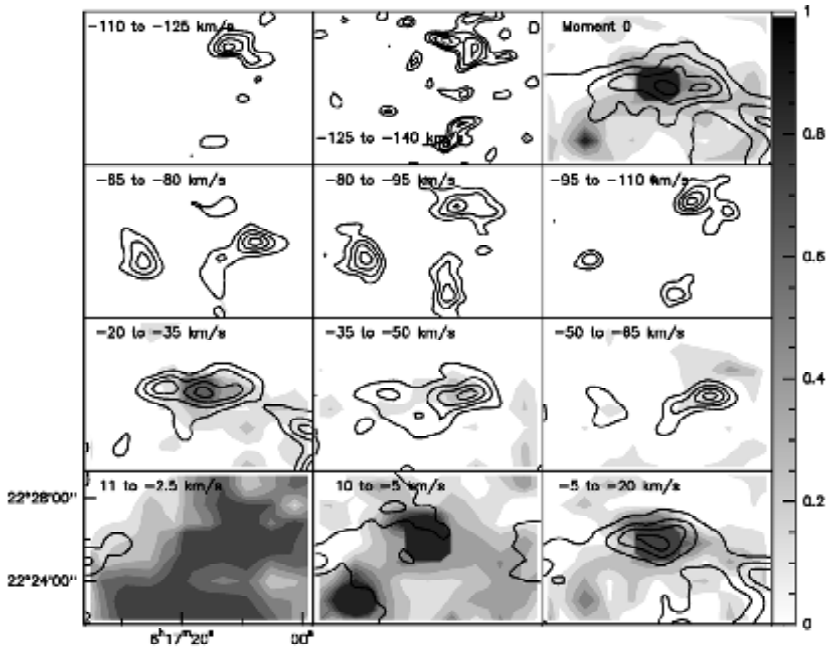

**Fig. 11.** Channel maps of $^{12}$CO and HI gas in clump B with the velocity range from -140 kms$^{-1}$ to 11 kms$^{-1}$. This region is also shown in the black box in Fig 6. The contours are HI gas, and grayscales are $^{12}$CO intensities. The contour levels are from 30% to 90% of the peak intensity, with a step of 20%. The top right panel shows the total integrated intensity of this region.

Fig 11 shows that, in a velocity channel width of 15 kms$^{-1}$, the channel maps of $^{12}$CO and HI gas in clump B with the velocity from -140 km s$^{-1}$ to 11 km s$^{-1}$. In the velocity range from -140 km s$^{-1}$ to -65 kms$^{-1}$, there are three high speed HI clumps, enclosing the central clump B region which is mainly concentrated in the velocity range from -50 to 10 km s$^{-1}$. In the velocity range from -140 to -65 kms$^{-1}$, basically no $^{12}$CO emission is detected, whereas strong $^{12}$CO emission is detected in the velocity range from -50 to 10 kms$^{-1}$ only in the central clump B region. The kinematics and distribution of cold (CO and HI) gas in clump B illustrate the possible explanation for the existence of multiple shocks on such a small physical scale as detailed in the next sub-section.

**4.2. Discussion of the multi-shock Models**

Though several shock tracers have been observed in IC 443, such as 21 cm HI spectra, H$_2$ rotational transition, and several other molecular lines[33][30][27], these observations can not be explained with a single velocity shock model. The following are several popular models used to explain the observations.

First, Burton et al. (1988)[4] proposed a radiative cooling mechanism after J-type shocks to explain the H$_2$ ratios. Molecular gas is dissociated and reforms after a J-type shock, which makes mixture of different velocities and temperatures. But, this model is only valid in the case of fast J-type shock. Given a slow J-type shock, a significant curvature should be derived in a rotational diagram from theoretical estimation, but in fact the curve is almost independent on the shock speed[19].

Second, Smith et al. (1991), and Neufeld & Yuan (2008) proposed a bow shock model. The bow shock model can well explain the H$_2$ observations, but it is not natural for small regions like clump B. With a diameter of 10 pc, IC 443 presents a clear and complete shell structure in such a large region. Under this model, it is hard to explain the coexistence of different velocity components in such small regions of 1-2 pc. In one hand, since the shell structure has been propagating for several thousand years, and it still keeps symmetry, while the anisotropism of bow shocks can easily destroy the structure. In another hand, the shock interface is almost planar on large scales, it is not reasonable to have many small bow shocks there.

Third, Wang & Scoville (1992, hereafter WS92) used a molecular clouds collision model as the explanation[31]. They used the interferometry observation of $^{12}$CO (J=2-1) in clump C to investigate the shock interaction with molecular clouds. Since they only used the velocity deviation between different sub-clumps of molecular gas, their model fits only shocks with a velocity below 50 km s$^{-1}$.

Based on WS92, we here propose a new model to explain the existence of multiple shocks on such a small physical scales of ~ 1-2 pc in IC 443. Here we take the clump B as an example to explain the origin of the multi-phase shocks.

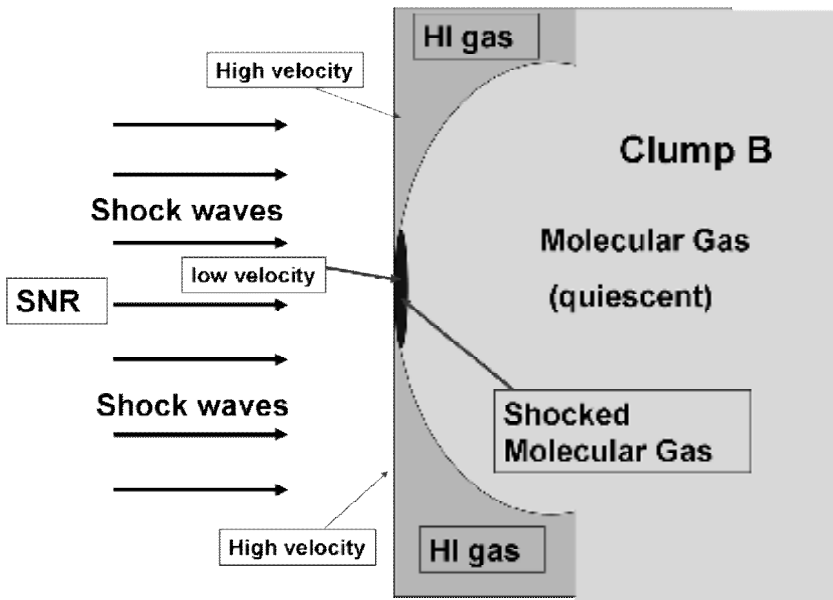

**Fig. 12.** The schematic diagram of the interaction between the shock and both the molecular and the atomic gas in clump B (the black area marked as "shocked molecular gas").

Fig 12 shows the schematic diagram of the shock collision with molecular clouds. We suppose that the molecular gas is not initially uniformly distributed, and clump B extends slightly towards the shock waves of SNR. The molecular gas in clump B has more or less a spherical structure, surrounded by atomic HI gas (Fig. 11) which plays an important role in the inter-clump medium. When the planar shock impacts the molecular clouds, the molecular clump B in the central region of HI will first interact with the shock. In this region, the shock could be of a slow J-type, because most of the shocked gas has a velocity below 60 km s$^{-1}$ (Figs. 7 & 11). Whereas in the outskirt regions, the shock interacts with atomic HI gas with lower density, so the shock can remain on a higher speed of 60 km s$^{-1}$ ≤Vs <100 km s$^{-1}$.

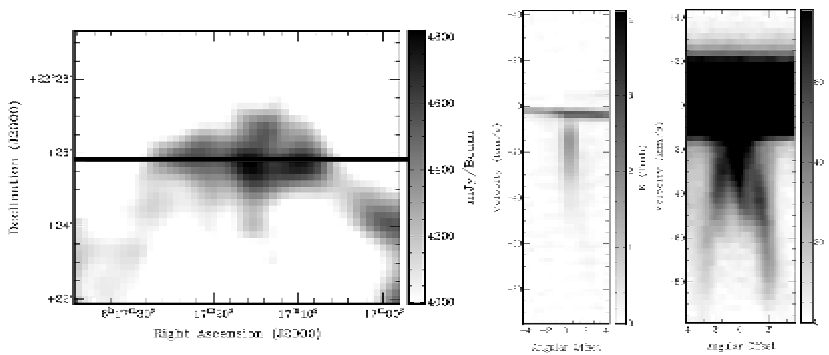

**Fig. 13.** The left panel shows the integrated intensity of HI gas in the velocity range from -100 km s$^{-1}$ to -10 kms$^{-1}$. The bold line shows the cut for the Position-Velocity (P-V) diagrams along the RA direction. The central panel shows the P-V diagram of $^{12}$CO. The right panel shows the P-V diagram of HI gas at the same position.

The following part examines our new model with the Position-Velocity (P-V) diagram analysis of both $^{12}$CO and HI. In the moment zero image map of shocked HI (left panel of Fig 13.), extended HI clumps surround the strongest clump B in the central region. The central panel is the P-V diagram of $^{12}$CO. The $^{12}$CO emission in the velocity range from 0 km s$^{-1}$ to -5 km s$^{-1}$ comes from ambient molecular gas, whereas $^{12}$CO has strong emission in the velocity range from - 5 km s$^{-1}$ to - 50 km s$^{-1}$ only in the offset position of 0′(centre of clump B), presenting a single peak profile. The right panel shows the HI P-V diagram, which shows extended surrounding emission in the velocity from 20 km s$^{-1}$ to -20 km s$^{-1}$. In the velocity range from - 20 km s$^{-1}$ to - 100 k ms$^{-1}$, the P-V diagram shows a trident shape in the offsets of -2′, +2′ and 0′ separately. The highest velocities of HI gas in the offsets of -2′ and +2′ reach nearly -100 km s$^{-1}$, whereas the peak HI emission in the offset of 0′ has a much lower velocity (~ - 20 km s$^{-1}$ to - 60 km s$^{-1}$), which is consistent with the velocity coverage of the $^{12}$CO emission peak. As the shock slows down, the HI emission clumps in the offsets of -2′ and +2′ move closer and closer to the central region. And finally, when the velocity slows down to -5 km s$^{-1}$, all these three emission peaks mix up, making it difficult to distinguish from the environmental gas.

The remarkable difference in P-V diagrams between $^{12}$CO and HI presents the diverse shock structures in different positions. In the offset of -2′ and +2′, the fast shock interacts with HI gas at a speed of ~100 kms$^{-1}$, while in the offset of 0′, the shock interacts with molecular gas, and slows down to a velocity of <-60 kms$^{-1}$. When the shock speed is higher than 20 km s$^{-1}$, H$_2$ molecular gas can be dissociated into HI atom, and when the shock speed is higher than 40-50 km s$^{-1}$, all the H$_2$ molecular gas could be dissociated. So from all of these various indications, we propose the following scheme. In the central region with the offset of 0′, HI gas comes from the dissociation of molecular hydrogen by a slow J-shock, while in the positions with the offsets of -2′ and +2′, the shock impacts the HI gas directly and thus keeps a relevant faster velocity of ~ 100 km s$^{-1}$.

As an independent check, we calculate the timescale of the shock propagation. We infer that the shock has a uniform initial speed of 100 km s$^{-1}$. Assuming a semi-sphere morphology with a 0.5 pc radius for the molecular cloud clump B, the timescale for the shock to go through the radius is

$$\tau = \frac{R}{v}, \quad \text{with} \quad R = 0.5 \text{ pc}, \quad v = 100 \text{ km s}^{-1}$$

The time scale $\tau$ is 5x10$^4$ years, which is fairly long enough to keep this observed morphology. Thus our model is also reasonable in terms of the time scale estimation.

### 4.3. The dissociation rate of the shocked molecular hydrogen

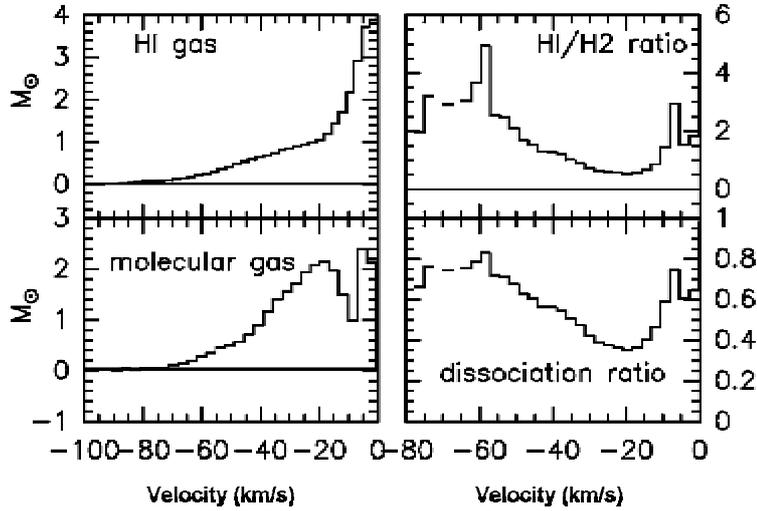

**Fig. 14.** The top-and bottom-left panels show the mass of HI gas and molecular gas as a function of velocity, respectively. Each bin has a velocity width of 3.5 km s$^{-1}$. The top right panel shows the mass ratio between the HI gas and molecular gas, and the bottom right panel shows the dissociation ratio, as a function of velocity. In the velocity range of < -60 kms$^{-1}$, we only plotted the ratio with 3δ data.

Fig 14 presents the mass distribution of HI and molecular gas as a function of velocity in the 1 square arcmin region of central clump B. Assuming $^{12}$CO is optically thin, adopting the excitation temperature of 80K, we use Equation 7 in Snell et al.(1984)[25] to estimate the total CO column density. When the velocity is <60 km s$^{-1}$, the signal to noise ratio is too low to get the ratios. When the velocity is >-20 kms$^{-1}$, HI emission is dominated by environmental gas. So only the mass ratio between the HI gas and molecular gas in the velocity range from - 60 km s$^{-1}$ to - 20 km s$^{-1}$ has enough confidence level. The dissociation ratio a=$M_{HI}$/($M_{HI}$+$M_{H2}$) changes from 35% in the velocity of -20 km s$^{-1}$ to 80% in the velocity of -60 km s$^{-1}$. The greater the shock speed propagates, the higher the dissociation rate achieves.

Uncertainty still remains. For example, gas accumulated by shock may contribute much to HI emission; The optically thin assumption also renders the molecular mass uncertainty. Uncertain filling factor can be crucial in the calculation too. So we have not fitted the correlation between the dissociation rate and shock speed, but only clarified the trend of dissociation changed by shock speed. The scarcity of HI gas in the velocity range of less than - 60 kms$^{-1}$ supports our former assumption, that is, HI gas mainly comes from disassociation from $H_2$ molecule in the central clump B.

## 5. The shocked spectra of $^{12}$CO and $^{13}$CO in clump B

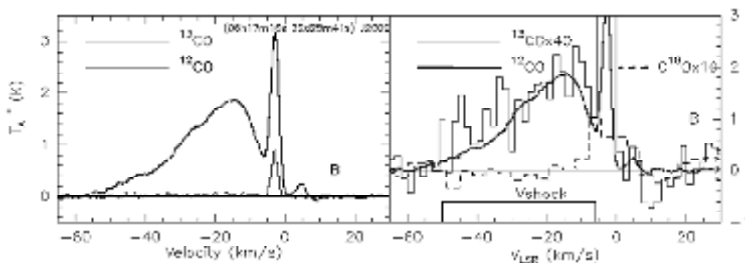

**Fig. 15.** $^{12}$CO (thick),$^{13}$CO (thin) and C$^{18}$O (dashed) spectra toward the centre B clump in IC 443. The left panel

shows the original overlaid spectra. The right panel shows the same spectra, but $^{13}$CO and C$^{18}$O spectra are multiplied by factors of 40 and 10 respectively.

Fig 15 presents the long time deep integration CO isotopic spectra in the centre of clump B. The most prominent linewing of $^{12}$CO much broadened line component (-60 km s$^{-1}$ to -6 km s$^{-1}$) has very weak $^{13}$CO counterpart. The optically thin lines like $^{13}$CO are less affected in the line profile by the shock from the SNR.

For the sake of the comparison with $^{12}$CO , we have amplified the $^{13}$CO spectra to an order of 40. As shown in the right panel of Fig 15, $^{13}$CO spectral line profile matches $^{12}$CO profile fairly well in the velocity range from -60 km s$^{-1}$ to -6 km s$^{-1}$. The ratio of integrated intensity of $^{12}$CO/$^{13}$CO in this velocity range is 38±2, which is smaller than the terrain isotopic ratio, but much larger than the ratio in normal molecular clouds (typically 5-10, as seen in the left panel in Fig. 15, the $^{12}$CO and $^{13}$CO line profile peaks around -3 km s$^{-1}$). Apparently, the physical processes involved in the shock produced this largely enhanced $^{12}$CO/$^{13}$CO ratio since the quiescent component unaffected by the shock at the same position has a normal $^{12}$CO/$^{13}$CO ratio usually found in typical molecular clouds (Zhang & Gao 2010, in preparation).

We here present two possible scenarios as explanations.

1) In the velocity range from -60km s$^{-1}$ to -6 km s$^{-1}$, $^{12}$CO spectra are optically thin, $\tau \ll 1$. Then the isotopic ratio of [$^{12}$C/$^{13}$C] is ~40 for the shocked gas. This would be one of the best methods to get the isotopic ratio of Carbon in such shocked regions. But it conflicts with the results from other observations[32],

2) Assuming that the $^{12}$CO spectra in this velocity range have a mediate optical depth, we can get the optical depth if the isotopic ratio of [$^{12}$C/$^{13}$C] is known. Then we can get a rather precise column density with both the line ratio and isotopic ratio.

We take the latter possibility and the results on the isotopic ratio of [$^{12}$C/$^{13}$C] in the literature. White et al.(1994) found the isotopic ratio of [$^{12}$C/$^{13}$C] in clump C to be near 80[32], which is derived from the observations of $^{12}$CO (J=2-1) and $^{13}$CO (J=2-1) with JCMT. So, if we take the similar isotopic ratio of 80 in clump B, our shocked $^{12}$CO spectra will have an optical depth of ~0.7, close to unity. This result contradicts with that in D92, which shows the optical depth as $\tau \ll 1$.

After carefully examining our CO line ratios observed in the calibrator, the pointing accuracy, and main beam size etc., we believe that our result of the intermediate optical depth in clump B is much favored as detailed in the following:

<1> Position switching mode was used both in our observation and in D92. The reference point in D92 was located at a long distance to clump B, which gave bad baselines and standing waves, as was mentioned in their paper, while our reference point is much closer(~10′) to clump B, which could result in more stable spectra.

<2> In D92, the $^{12}$CO and $^{13}$CO spectra were observed separately, so the pointing drifts as well as calibrations between $^{12}$CO and $^{13}$CO observations would be amplified in the final derived ratios. Whereas in our observations, the $^{12}$CO and $^{13}$CO spectra are observed simultaneously, which can perfectly avoid these problems.

<3> The mainbeam size (FWHP) in D92 was 46″, while in our observation it is $50''\pm 7$ in the direction of azimuth, and $54''\pm 3$ in the direction of Elevation. Since the source of clump B is only slightly larger than the beamsizes, these two telescopes have different filling factors and our slightly larger telescope beam is preferred.

We compare the ratio of peak intensity of $^{12}$CO in quiescent gas component and in shocked gas component for both observations. This ratio is ~1.5 in D92, but ~2 in our result, which corroborates the different beam filling factor effect.

To conclude, the optical depth is calculated with the intensity ratio of $I_{^{12}CO}/I_{^{13}CO}$ in the shock influenced velocity range, which is not much affected by filling factor. The simultaneous observations of both $^{12}$CO and $^{13}$CO avoid the uncertainties of the pointing errors and different calibrations. At the same time, our adopted much closer reference position enhances the quality of spectra, which generates better and more reliable observational results.

## 6. Conclusion

We have made a large scale, high spatial sampling, and simultaneous $^{12}$CO and $^{13}$CO mapping observation of the SNR IC 443 and its molecular clouds. The multi-band images including optical, infrared, radio, X-ray etc. are compared with $^{12}$CO and $^{13}$CO morphologies. The kinematics of $^{13}$CO, as the quiescent gas tracer, and $^{12}$CO, as the shock influenced gas tracer as well as that of diffuse HI are examined and compared. In clump B, a new model is proposed to naturally explain the existence of multiple phase shocks in small regions. The dissociation ratio of shocked molecular gas is given based upon the derived ratio of molecular and atomic gas mass in clump B. Finally, we calculate the optical depth with the long time integration of $^{12}$CO and $^{13}$CO spectra in the shock influenced velocity range, and discuss the validity of the assumption of optically thin emission for the shocked $^{12}$CO.

*Acknowledgements: We are grateful to all staff of the 13.7m telescope of Purple Mountain Observatory for their dedicated assistance. We thank Dr. Lee, J.J. for providing the HI data and also thank Prof. Zheng, Xianzhong and Dr. Zhao, Yinghe for their constructive suggestion and advice. Research for this project is supported by NSF of China (Distinguished Young Scholars #10425313, grants #10833006, �), Chinese Academy of Sciences' Hundred Talent Program, and 973 project of the Ministry of Sci. and Tech. of China (grant #2007CB815406).*